\documentclass{article}

\newcommand{\lang}{brazil}

\usepackage{sbc/sbc}
\usepackage[\lang]{babel}
\usepackage{graphicx,url}
\usepackage[utf8]{inputenc}
\usepackage[T1]{fontenc}
\usepackage{subfigure}
\usepackage{caption}
\usepackage{hyperref}
\usepackage{booktabs}
\usepackage{multirow}
\usepackage{booktabs}
\usepackage{tipa}

\author{
    Alan Veloso\inst{1},
    Jeffson Sousa\inst{1, 2},
    Bruno Evaristo\inst{2},
    Antônio Abelém\inst{1}
}

\graphicspath{{i18n/\lang/fig/}}

\begin{document}

\pagestyle{myheadings}


\title{MinIndy: Uma Ferramenta de Início Rápido do Hyperledger Indy\footnote{Este trabalho é parte do projeto aprovado no Programa de Gestão de Identidade 2023 da Rede Nacional de Ensino e Pesquisa (RNP).}}

\address{
    Grupo de Pesquisa em Rede de Comutadores e Comunicação Multimídia (GERCOM)\\
    Universidade Federal do Pará (UFPA)\\
    Belém -- PA -- Brasil
    \nextinstitute
        Centro de Pesquisa e Desenvolvimento em Telecomunicações (CPQD)\\
        Campinas -- SP -- Brasil
    \email{
        aveloso@ufpa.br,
        \{jcsousa, elderb\}@cpqd.com.br,
        abelem@ufpa.br
    }
    \vspace{-0.2cm}
}

\maketitle

\begin{resumo}

A plataforma blockchain Hyperledger Indy, voltada para redes de gestão de identidade, tem ganhado importância, mas a instanciação de uma rede completa é complexa e exige experiência. Portanto, o presente trabalho descreve o MinIndy, uma ferramenta projetada para simplificar a instalação e a configuração de redes Hyperledger Indy. Essa simplificação permitirá que pessoas com menor nível de expertise possam criar suas redes Indy. O que a torna uma alternativa viável para organizações que buscam adotar redes Blockchain Hyperledger Indy com menor esforço.

\end{resumo}


\section{Introdução}
\label{sec:intro}

A tecnologia Blockchain tem em seu leque de soluções a Hyperledger Indy, focada em redes de gestão de identidade e mantida pela Hyperledger Foundation \cite{monrat2019survey}. Essa solução é utilizada pela Sovrin\footnote{\url{https://sovrin.org/}}, uma das redes permissionada mais conhecidas, que conta com a participação e interesse de diversas organizações. Participar dessa e de outras redes similares oferece benefícios às organizações, como redução de intermediários, custos, e maior segurança e privacidade dos dados. Para usufruir desses benefícios, as organizações precisam instanciar e configurar um nó na rede, tarefa considerada não trivial, extensa e propensa a erros até para profissionais experientes \cite{silva2022relatos}. Isso exige tempo e conhecimento especializado, o que pode impedir a participação de algumas organizações.

Para simplificar o processo de instanciação de uma rede blockchain existem iniciativas, como o Minifabric\footnote{\url{https://labs.hyperledger.org/labs/minifabric.html}} e o Microfab\footnote{\url{https://labs.hyperledger.org/labs/microfab.html}}. Contudo, elas são voltadas para redes Fabric de propósito geral. Até o momento, não há iniciativas de simplificação para a Indy mantidas pela Hyperledger Labs\footnote{\url{https://labs.hyperledger.org/}}, responsável pela curadoria de projetos Hyperledger.

Este trabalho propõe uma solução para essa lacuna, objetivando simplificar a instanciação de uma rede Indy por meio da automatização do processo de instanciação e configuração dos nós. A automatização possibilita reduzir ou eliminar algumas etapas, minimizando tempo e evitando possíveis erros, pois os parâmetros de automatização estarão predefinidos correntemente.

O restante do trabalho está organizado da seguinte forma: a metodologia utilizada para o desenvolvimento é apresentada na \autoref{sec:gott}; a proposta de ferramenta é detalhada na \autoref{sec:propose}; por fim, a \autoref{sec:con} descreve as conclusões gerais.


\section{Metodologia}
\label{sec:gott}

A metodologia utilizada neste trabalho foi dividida em duas partes: (i) definição dos principais processos para a instalação e a configuração de uma rede blockchain Hyperledger Indy; (ii) desenvolvimento da ferramenta para automatizar esses processos.

Para definir os principais processos para a instalação e a configuração de uma rede blockchain Hyperledger Indy, foram coletadas e sumarizadas as informações contidas na documentação da Hyperledger Foundation. Essa abordagem garante que os processos identificados sejam baseados nas melhores práticas e orientações estabelecidas pela comunidade de desenvolvimento do Hyperledger Indy.

Para automatizar esses processos, serão construídos \textit{scripts}. Esses \textit{scripts} automatizarão a instalação e a configuração dos principais passos para instanciar um nó Hyperledger Indy, possibilitando adequar diversos aspectos relacionados ao funcionamento e segurança da rede.

\section{MinIndy}
\label{sec:propose}

Esta seção descreve o MinIndy e informações importantes para o seu desenvolvimento. Será apresentada uma visão geral de uma rede Indy, com os integrantes, componentes, e os processos identificados como os principais para a instanciação de uma rede Indy baseado na documentação do Indy \cite{hyperledgersetting}. Esses processos serão automatizados pelo MinIndy. Também serão apresentadas as configurações padrão utilizadas para automatização. A seguir é apresentada a arquitetura da solução, bem como as ferramentas de automação utilizadas no desenvolvimento do MinIndy.

Os integrantes das redes Indy podem ter dois papéis, \textit{Trustee} e \textit{Steward}: os \textit{Trustee} são as pessoas responsáveis por gerenciar a rede e proteger a integridade da governança da rede. Em uma rede de produção, é necessário ter, no mínimo, três \textit{Trustees}, ou seja, três pessoas diferentes são obrigatórias, mas é desejável ter mais; um \textit{Steward} é uma organização responsável por manter um nó da rede. Inicialmente, quatro \textit{Stewards} são fundamentais para estabelecer uma nova rede, o que significa que são necessárias quatro organizações diferentes. Mais \textit{Stewards} podem ser adicionados posteriormente. 

Os \textit{Stewards} operam os nós validadores (\textit{Validator}) da rede. O \textit{Validator} é o principal componente que forma uma rede Indy, ele é a máquina que se tornará parte de uma rede Indy. Um \textit{Validador} permitirá que a organização faça parte do que é chamado de consenso. Por padrão, um \textit{Steward} só tem permissão para operar um \textit{Validador} por rede.

Os dois processos identificados como os principais para a instancionação de uma rede Indy são: a criação de uma rede e a adição de um novo nó. Esses processos são descritos de maneira geral a seguir:

\begin{itemize}
    \item Criação de uma Rede: A etapa de criação de uma rede Indy envolve a criação de informações de identificação criptográficas dos \textit{Trustees} e \textit{Stewards}. Além disso, os \textit{Stewards} também devem definir e criar informações de rede e criptográficas dos seus \textit{Validators} como, endereço IP, chave pública BLS (Boneh–Lynn–Shacham) e outros. Essas informações são utilizadas para definir os administradores e os nós confiáveis iniciais, respectivamente. Após todas as informações e outras informações gerais da rede terem sido definidas e compartilhadas entre os \textit{Validators}, é possível inicializar a rede.
   
    \item  Adição de um Novo Nó: Para adicionar um novo nó é necessário que um administrador da rede, ou seja, um \textit{Trustee}, adicione um novo \textit{Steward}, para que este tenha a permissão de adicionar um \textit{Validator}. Após o \textit{Steward} ter definido as informações de rede e de identificação necessários, ele usará os arquivos de configuração da rede para iniciar o seu \textit{Validator}. Por fim, o \textit{Steward} adicionará o seu nó à rede.
\end{itemize}

A instanciação de uma rede Indy possibilita a personalização de algumas configurações. Para está proposta, serão utilizadas as configurações da rede Sovrin. As configurações da Sovrin foram escolhidas por serem as mais difundidas entre as iniciativas que utilizam a tecnologia Indy. Além disso, é a rede cujo desenvolvimento conta com o maior número de organizações contribuintes, o que torna as configurações utilizadas mais atraentes para a maioria dos interessados. Assim, ao automatizar as configurações dessa rede, possibilita que outras organizações façam parte da rede Sovrin e também construam suas próprias redes baseadas nessas configurações.

\begin{figure}[!htb]
    \centering
    \includegraphics[width=.75\textwidth]{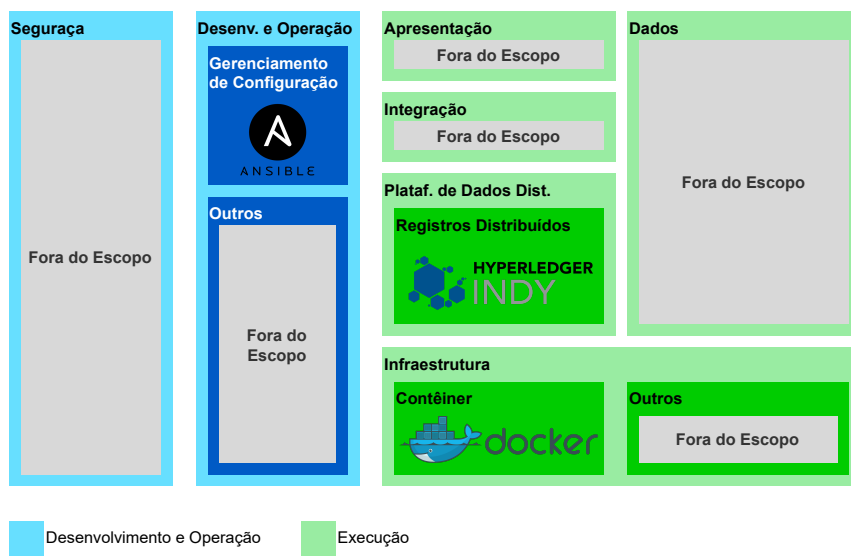} \\
    \caption{Arquitetura do MinIndy}
    \label{fig:arch_minidy}
\end{figure}

A arquitetura de referência \cite{hyperledgerarchitecture}, fornecida pela Hyperledger Foundation, serviu de base para o desenvolvimento da arquitetura do MinIndy apresentada na \autoref{fig:arch_minidy}. O MinIndy abarca os níveis de desenvolvimento e operação, plataformas de dados distribuídos e infraestrutura. Os demais níveis estão fora do escopo, por já serem tratados por aplicações de escolha do usuário ou por adicionarem complexidade à solução.

Os recursos de \textit{desenvolvimento e operação} permitem desenvolver serviços dentro ou fora do registro distribuído e baseados em SDKs e IDEs (por exemplo, APIs da Web). Também pode se tratar da manutenção, do monitoramento e da administração do registro distribuído e seus serviços. O MinIndy atua especificamente no gerenciamento de configuração, que envolve a automação de \textit{scripts} e permite que os operadores configurem redes blockchain de forma similar, usando alterações mínimas de configuração. Para isso, o MinIndy usa o Ansible, uma ferramenta de automação de TI que simplifica tarefas de gestão de infraestrutura. Por ser compatível com vários sistemas operacionais, o Ansible permite criar \textit{scripts} para aumentar a eficiência e a produtividade do gerenciamento de infraestrutura.

A \textit{plataforma de dados distribuídos} forma o núcleo da solução. O MinIndy dá suporte à blockchain Hyperledger Indy. No nível de \textit{infraestrutura}, estão funcionalidades necessárias para executar ou implantar os diferentes serviços de uma arquitetura blockchain. Os recursos de \textit{contêiner} permitem que os usuários implantem e gerenciem a rede Indy usando virtualização baseada em contêineres. Eles também permitem que um desenvolvedor empacote e envie um aplicativo com todas as partes necessárias, como bibliotecas e outras dependências. Optou-se pelo Docker, o que garante um ambiente uniforme e consistente de execução e possibilita a instalação e a execução em ambientes compatíveis com o Docker, independentemente do sistema operacional ou da plataforma de nuvem.


\section{Conclusão} 
\label{sec:con}

Este artigo apresentou a ferramenta MinIndy, ainda em desenvolvimento, que visa simplificar o processo de instanciação e configuração de uma rede Indy. Para essa automatização, o projeto utiliza a plataforma de automatização, Ansible, e de virtualização, Docker.

A ferramenta MinIndy pode ser útil, tanto para o desenvolvimento quanto para a pesquisa. Para o desenvolvimento, a automatização dos processos de instanciação reduzirá o esforço necessário para a criação de uma rede ou nó Indy, permitindo que outros indivíduos ou organizações participem da rede, mesmo sem ter o conhecimento ou a disponibilidade para adquiri-los. Para a pesquisa, a ferramenta fornecerá uma alternativa para os pesquisadores realizarem estudos relacionados ao Indy ou suas aplicações sem precisar configurar uma rede Indy, ou ter acesso a uma rede já existente.


\section*{Agradecimentos}

Este trabalho foi realizado com o apoio do Conselho Nacional de Desenvolvimento Científico e Tecnológico (CNPq) e da Rede Nacional de Ensino e Pesquisa (RNP).



\bibliography{references}

\end{document}